\documentstyle[prd,aps]{revtex}
\begin{document}
\draft
\input epsf
\twocolumn[\hsize\textwidth\columnwidth\hsize\csname
@twocolumnfalse\endcsname

\title{Gauge Quintessence}
\author{L. Pilo$^{(1)}$, D.A.J. Rayner$^{(2)}$, A. Riotto$^{(2)}$}

\address{$^{(1)}${\it Service de Physique The\'orique,
CEA/Saclay, Gif-sur-Yvette, France}}
\address{$^{(2)}${\it Department of Physics and INFN, Sezione di Padova, 
via Marzolo 8, I-35131
Padova, Italy}}

\date{February, 2003}
\maketitle
\begin{abstract}
We discuss a new model of quintessence in which the quintessence field
is identified with the extra-component of a gauge field
in a compactified five-dimensional theory. We show that
the extremely tiny energy scale $\sim (3\times 10^{-3}\:{\rm eV})^4$ 
needed to account for the present
acceleration of the Universe can be naturally explained
in terms of high energy scales such as the scale of Grand Unification.

\end{abstract}
\vskip2pc]

\def\simlt{\stackrel{<}{{}_\sim}}
\def\simgt{\stackrel{>}{{}_\sim}}

There is an increasing evidence that the energy density of
(baryonic plus dark) matter   in the Universe is smaller
than the critical density \cite{reviewparameters}.
If the Universe is flat, as predicted by
the most natural inflation models \cite{reviewinf} and confirmed
by the recent measurements of the cosmic microwave background 
anisotropies \cite{reviewcmb} , an  
additional
dark energy density is necessary    to reach $\Omega_0=1$. 
This dark energy seems to be the predominant form of energy in the
present Universe, about $70\%$ of the critical energy density, 
and should possess a negative pressure $p$.
An obvious candidate is represented
by the cosmological constant, whose equation of state
is $\rho=-p$. If this is the
option chosen by Nature, particle physicists have to face the Herculean
task to explain why the energy of
the vacuum $V_0$ is of the order of $(3\times 10^{-3}\:{\rm eV})^4$.
Another possibility invokes a mixture of
cold dark matter and quintessence \cite{quint}, a
slowly-varying, spatially inhomogeneous component
with equation of state $p_Q=w_Q\rho_Q$, with $-1< w_Q\leq 0$.
The role of  quintessence  may be played by any  scalar field $Q$
which is slowly rolling down its potential $V(Q)$.
The slow evolution is needed to obtain a negative
pressure, $p_Q=\frac{1}{2}\dot{Q}^2-V(Q)$, so that
the kinetic energy density is less than the potential energy density.
The quintessence 
field $Q$ rolls down a potential
according to  the equation of motion $
\ddot{Q}+3H\dot{Q}+V'(Q)=0$, 
where $H$ is the Hubble constant satisfying the Friedmann equation
in a flat Universe 
\begin{equation}
H^2=\left(\frac{\dot{a}}{a}\right)^2=\frac{1}{3 M_p^2}\left(
\frac{1}{2}\dot{Q}^2+V(Q)+\rho_{B}\right), 
\end{equation}
where $a$ is the scale factor, $M_p=2\times 10^{18}$ GeV is the reduced
Planck scale and $\rho_B$ is the remaining background energy density.

Since at present 
 the quintessence field $Q$ dominates the energy density
of the Universe, one can write $\frac{1}{2}\dot{Q}^2=\frac{3}{2}\left(
1+w_Q\right)H^2M_p^2$ and $V(Q)=\frac{3}{2}\left(
1-w_Q\right)H^2M_p^2$. 
Let us assume that  the  quintessence potential has the
parametric form 
\begin{equation}
V(Q)=V_0 \,\,\,{\cal V}\left(\frac{Q}{f}\right)\, ,
\label{form}
\end{equation}
where $V_0$ parametrizes the height of the potential.
Let us borrow the notation traditionally adopted in inflation model-building
by defining a slow-roll parameter $\epsilon=-\dot{H}/H^2$.
If the Universe is suffering an acceleration stage
because of the quintessence dynamics, then $\ddot{a}/a=
\left(1-\epsilon\right)H^2>0$ and the 
parameter $\epsilon\sim
M_p^2\left(V^\prime/V\right)^2\sim \left(M_p/f\right)^2$ 
has to be smaller than   unity. This implies
that the scale $f$ has to be larger  than   the Planck scale. 
In turn, the quintessence 
mass $m_Q$ must be extremely tiny since 

\begin{equation}
m_Q^2\sim V^{\prime\prime}
\sim \frac{V_0}{f^2}\sim H^2\left(\frac{M_p}{f}\right)^2\, .
\end{equation}
 The quintessence field has to roll
down its potential with a mass comparable (or smaller) than 
$H\sim 10^{-42}$ GeV!

The extreme flatness of the quintessence field represents
a real challenge from the particle physics point of view and there
are no completely natural models of quintessence.
Supersymmetry is usually invoked to preserve the potential
from acquiring large corrections to the  mass of the quintessence
field. However, the flatness of the potential tends to be  spoiled when
supergravity \cite{sugra} corrections are included \cite{sugraproblems}. 
The same
problem manifests itself in trying to build-up a satisfactory 
model of inflation \cite{reviewinf}.

Another possibility is to consider the quintessence field as 
a pseudo Nambu-Goldstone boson (PNGB) \cite{josh}, {\it i.e.}
the underlying theory possesses a non-linearly realized
symmetry and the quintessence field can be parametrized through
an angular variable $\theta=Q/f$. In the limit of exact symmetry
the quintessence field $Q$ does not have a potential which is generated only
in the presence of an explicit breaking term. The same effect could explain
why the quintessence field does couple to ordinary matter more weakly than
gravity.

If the quintessence field is a PNGB, its Lagrangian
can be written as 

\begin{equation}
{\cal L}=\frac{1}{2}
\left(\partial_\mu Q\right)^2-V_0\left[1-\cos\left(\frac{Q}{f}\right)\right]
\, ,
\label{e1}
\end{equation}
where $f$ is the spontaneous breaking scale. 

The problem of identifying the quintessence field with a 
PNGB comes from the fact that the spontaneous breaking scale
$f$ has to be comparable to  the Planckian scale and, therefore, the
effective four-dimensional field theory description is expected to break down 
due to quantum gravity corrections.
One should note however that there might be 
shift symmetries, for example
acting on the model-independent axion, which constrain the form of these 
quantum corrections to be small in some regions of parameter space (see
Kim and Nilles in Ref. \cite{josh} and \cite{choi}). 
The existence of such a  symmetry is
not imposed from a four-dimensional  theory, but 
it is deduced from the string theory.

To summarize, there are two necessary key steps one needs to take in order
to build up a successful quintessence particle physics model. One is 
to explain the reason why the scale $V_0$ -- 
parametrizing  the height of the potential -- is so tiny and the other
is to explain 
why the overall scale $f$ spanned by the quintessence field 
may be comparable to the Planckian scale without running into
trouble with the four-dimensional description.
Motivated by similar recent considerations applied to  models
of primordial inflation
\cite{newwilson1,newwilson2,newwilson3},
in this short note
we would like to show that the extra-dimensional 
generalization of identifying the quintessence field with a PNGB
may help in taking both steps.

We 
consider a five-dimensional model with the extra fifth dimension 
compactified on a circle of radius $R$  and identify the
quintessence field with the fifth component $A_5$ of an abelian gauge field
$A_M$ ($M=0,1,2,3,5)$
propagating in the bulk (the generalization to the  non-abelian 
case is straightforward). As such, the quintessence field cannot have
a local potential because of the higher-dimensional gauge invariance. 
However, a non-local potential as a function of the 
gauge-invariant Wilson line

\begin{equation}
e^{i \theta}=e^{i\,\oint\,g_5 A_5\,dy}\, ,
\end{equation}
where $y$ is the coordinate along the fifth dimension, $0\leq y< 2\pi R$, 
will be generated in the presence of fields charged under the
abelian symmetry \cite{hosotani}

Writing the field $A_5$ as 
\begin{equation}
A_5=\frac{\theta}{2\pi g_5 R}\, ,
\end{equation}
where $g_5$ is the five-dimensional gauge coupling constant, at energies below
the scale $1/R$, $\theta$ looks like a four-dimensional field
with Lagrangian

\begin{equation}
{\cal L}=\frac{1}{2 g_4^2(2\pi R)^2}\left(\partial_\mu\theta\right)^2-
V(\theta)\, ,
\label{e2}
\end{equation}
where $g_4=g_5/(2\pi R)^{1/2}$ is the four-dimensional gauge coupling 
constant. Comparing Eqs. (\ref{e1}) and (\ref{e2})
one identifies the overall scale $f=\frac{1}{2\pi g_4 R}$ and the
field $Q=\frac{\theta}{2\pi g_4 R}$.
Therefore, one can easily see that the overall scale $f$ may be 
comparable
to the four-dimensional Planckian scale, $f\sim M_p$, if the
four-dimensional
constant is small enough \cite{newwilson1,newwilson3}. 
For instance, requiring that $1/R\sim 10^{16}$ GeV
imposes that $g_4\sim 10^{-3}$.  The higher-dimensional nature
of the theory preserves the quintessence potential from acquiring 
dangerous corrections, and non-local effects must be
necessarily exponentially suppressed because the typical length
of five-dimensional quantum gravity effects $\sim M_5^{-1}$, where
$M_5$ is the five-dimensional Planck scale, is much smaller than the
size of the extra-dimensions. 

Let us now turn to the form of the potential. We assume that the 
potential for the quintessence field is generated radiatively by a set
of bulk fields which are charged under the $U(1)$ symmetry with charges
$q_a$. The fundamental hypothesis we make is that these bulk fields
possess a bare mass $M_a\gg R^{-1}$ and that 
there is no charged matter with mass
below the compactification scale.
The masses $M_a$ may be generated by some
gauge symmetry breaking phenomenon at scales larger than $R^{-1}$.
For instance, bulk fields may be charged under another abelian factor
broken at energies larger than the compactification scale.
From the 
four-dimensional point of view, this is equivalent to having  a 
tower of Kaluza-Klein states with squared masses 

\begin{equation}
m_a^2=M_a^2+\left(\frac{n}{R}+g_4\, q_a\, Q\right)^2\, ,\,\,\,
\,\, (n=0,\pm1,\pm2,\dots)\, .
\end{equation}
Borrowing from finite temperature
field theory calculations, the $Q$-dependent part of the potential can be
written as \cite{pq}

\begin{equation}
V(Q)=\frac{1}{128\pi^6 R^4}\,{\rm Tr}\left[V\left(r^F_a,Q\right)
-V\left(r^B_a,Q\right)\right]\, ,
\end{equation}
where the trace is over the number of degrees of freedom, 
the superscripts $F$ and $B$ stand for fermions and bosons, 
respectively and
\begin{eqnarray}
\label{pot}
V\left(r_a,Q\right)&=&x_a^2\,{\rm Li}_3\left(r_a\,e^{-x_a}\right)+
3\,x_a\,{\rm Li}_4\left(r_a\,e^{-x_a}\right)\nonumber\\
&+&3\,{\rm Li}_5\left(r_a\,e^{-x_a}\right)+{\rm h.c.}\, .
\end{eqnarray}
We have defined 
\begin{eqnarray}
x_a&=&2\pi R M_a\, ,\nonumber\\
r_a&=& e^{\frac{i q_a Q}{f}}\, ,
\end{eqnarray}
and in Eq. (\ref{pot}) the functions ${\rm Li}_n(z)$ stand for   
the polylogarithm
functions 
\begin{equation}
{\rm Li}_n(z)=\sum_{k=1}^{\infty}\frac{z^k}{k^n}\, .
\end{equation}
The potential (\ref{pot}) shows many similarities with the potential
one obtains in four-dimensional field theories at finite temperature
where, 
in the imaginary time formalism,  four-dimensional loop integrals
become integrals over the three spatial momenta and a sum over
the so-called Matsubara frequencies. The finiteness of the
potential at finite temperature is due to the fact that particles
with wavelengths smaller than the inverse temperature 
have Boltzmann (exponentially) suppressed abundances in the plasma.
Similarly, the potential (\ref{pot})
is independent of any
ultraviolet cut-off. This is because the Wilson line is a global quantity
while ultraviolet effects are local. 

More crucial for  our
considerations is the behaviour of the potential when the bare masses
$M_a\gg R^{-1}$: the overall height of the potential is
exponentially suppressed! This can be easily understood by
thinking again of the four-dimensional finite temperature case. 
Particles in the plasma with bare masses $M$ much larger than the
temperature $T$ do not contribute to the effective potential
apart from tiny exponentially suppressed contributions. In the very same
way, bulk fields charged under the abelian gauge symmetry $U(1)$
give an exponentially suppressed contribution to the potential
(\ref{pot}) if their bare mass term is much larger than the
{\it effective} temperature $T=R^{-1}$.

Let us, for simplicity, assume that all bare masses $M_a$ are 
equal to a common mass $M\gg R^{-1}$. The potential (\ref{pot})
is well approximated by the form (\ref{form}) with

\begin{equation}
\label{crucial}
V_0\simeq \frac{c}{16\pi^4}\frac{M^2}{R^2}\,e^{-2\pi M R}=
\frac{c}{16\pi^4}\frac{M^2}{R^2}\,e^{-M M_p^2/M_5^3}\,,
\end{equation}
where $c={\cal O}(1)$ is a numerical coefficient depending upon the charges
of the bulk fields and in the last passage we have made use of the
relation $M_p^2=2\pi R M_5^3$.
We discover that the extreme
smallness of the  height of the potential
$V_0$ can be naturally explained with a moderate 
fine-tuning of the parameter $M R$. This is the main
result of this note.
To give a feeling for the numbers, 
setting $x_*=M_5/M_p$ and imposing the condition (\ref{crucial}) gives

\begin{equation}
\frac{M}{M_5}\simeq x_*^2\left(270+12\,{\rm ln}\,x_*\right)\, ,
\end{equation}
which fixes the parameter $MR$ to be

\begin{equation}
MR \simeq 40+2\,{\rm ln}\,x_*\, .
\end{equation}
If we do require that the overall scale $M$ is smaller than the
five-dimensional Planck scale $M_5$ in order to avoid 
dealing with loops of massive excitations of
quantum gravity, we have to require $M\ll M_5$. This corresponds to
a mild constraint on $M_5$, $M_5\ll 6\times 10^{-2}\,M_p$. 
The problem of explaining the smallness of the energy scale 
of the quintessence
potential is therefore exponentially reduced and requires only 
a moderate (logarithmic) fine-tuning.

The idea of identifying the
quintessence field with a Wilson line has been briefly discussed
and disregarded  in
Ref. \cite{newwilson1} which was devoted to propose  an
interesting  model of 
primordial inflation where the inflaton field is interpreted as 
the extra-component of a gauge field in a five-dimensional theory.
Ref.  \cite{newwilson1} considered
the case in which the charged bulk fields do not possess large bare masses.
The quintessence mass-squared $m_Q^2$ turns out to be  of the order of
$\left(f^2 R^4\right)^{-1}\sim g_4^2/R^2$ and an extreme
fine-tuning is needed either for the four-dimensional gauge coupling
 $g_4$ or for the radius of compactification $R$. Our findings
show that such a fine-tuning can be avoided.

In conclusion we have shown that the extra-component of a gauge field
in five-dimensions may be a good candidate for 
quintessence. The flatness of its potential is protected by
gauge invariance in the higher-dimensional world and the tiny scale of the
potential needed to accommodate the presently observed accelerating phase
of the Universe may be naturally obtained if the non-local 
potential for the quintessence fields is provided by massive
bulk fields. Our proposal 
does not solve however the so--called coincidence problem, that is why
the amount of dark energy density 
is of the same order as the energy density stored
in dark matter at the present epoch.

This work was supported in part by the RTN European Program 
HPRN-CT-2000-00148.

\end{document}